\documentclass[fleqn,usenatbib]{mnras}

\usepackage[T1]{fontenc}

\DeclareRobustCommand{\VAN}[3]{#2}
\let\VANthebibliography\thebibliography
\def\thebibliography{\DeclareRobustCommand{\VAN}[3]{##3}\VANthebibliography}

\usepackage{mathtools, amssymb, amsmath, glossaries, multirow, booktabs, comment, graphicx, lineno}

\usepackage{newtxtext,newtxmath}

\newcommand{\zs}{z_{\rm s}}

\newcommand{\rein}{\theta_{\rm Ein}}

\newcommand{\mbh}{M_{\rm BH}}

\newcommand{\msun}{M_{\odot}}
\newcommand{\kms}{km~s$^{-1}$}
\newcommand{\sfinf}{SF$_\infty$}
\newcommand{\stkout}[1]{\ifmmode\text{\sout{\ensuremath{#1}}}\else\sout{#1}\fi}

\newcommand{\rblr}{R_{\rm BLR}}

\defcitealias{OguriM+10}{OM10}
\defcitealias{Suberlak+21}{S21}

\title[Variability of Lensed QSOs]{Strong Lensed QSOs with Variability Detectable by LSST: How many are there?}

\author[Y. C. Taak and T. Treu]{
Yoon Chan Taak$^{1}$\thanks{E-mail: yctaak@astro.ucla.edu} and
Tommaso Treu$^{1}$
\\
$^{1}$Department of Physics and Astronomy, University of California, Los Angeles, CA 90095-1547, USA\\
}

\date{Accepted XXX. Received YYY; in original form ZZZ}

\pubyear{2023}

\begin{document}
\label{firstpage}
\pagerange{\pageref{firstpage}--\pageref{lastpage}}
\maketitle

\begin{abstract}
Strong lensed quasi-stellar objects (QSOs) are valuable probes of the universe in numerous aspects. Two of these applications, reverberation mapping and measuring time delays for determining cosmological parameters, require the source QSOs to be variable with sufficient amplitude. In this paper, we forecast the number of strong lensed QSOs with sufficient variability to be detected by the Vera C. Rubin Telescope Legacy Survey of Space and Time (LSST). The damped random walk model is employed to model the variability amplitude of lensed QSOs taken from a mock catalog by \cite{OguriM+10}. We expect 30--40\% of the mock lensed QSO sample, which corresponds to $\sim$1000, to exhibit variability detectable with LSST. A smaller subsample of 250 lensed QSOs will show larger variability of $>0.15$~mag for bright lensed images with $i<21$ mag, allowing for monitoring with smaller telescopes. We discuss systematic uncertainties in the prediction by considering alternative prescriptions for variability and mock lens catalog with respect to our fiducial model. Our study shows that a large-scale survey of lensed QSOs can be conducted for reverberation mapping and time delay measurements following up on LSST. 
\end{abstract}




\section{Introduction}

Quasi-stellar objects (QSOs) are one of the brightest objects in the universe, making them excellent probes to study the distant universe \citep{KimY+15,KimY+19,KimY+20,KimY+22,Banados+16,JeonY+16,JeonY+17,JiangL+16,WangF+21}. Their extreme luminosities, sometimes outshining their host galaxies, are powered by gas accretion into the central supermassive black hole (SMBH). 

The mass of the supermassive black hole ($\mbh$) at the QSO center is a crucial property used for understanding not only the history of SMBHs, but also their seeming coevolution with their host galaxies \citep{Ferrarese+00,Gebhardt+00,Kormendy+13}. Among the three main methods for measuring $\mbh$, the most reliable method uses the dynamics of gas clouds and stars rotating around the SMBH based on simple Newtonian dynamics \citep[e.g.,][]{Ghez+08,Cappellari+09}, but this is only feasible for local SMBHs for which the orbit can be precisely monitored. 

Reverberation mapping (RM), in which the sphere of influence of the black hole is resolved in time rather than in angles, is commonly used for more distant SMBHs \citep{Blandford+82,Fromerth+00,Peterson+04,ShenY+15b}. The broad line region (BLR) is excited by the continuum emitted from the accretion disk, so fluctuations in the continuum are mirrored in the emission line light curve with a time lag, which is interpreted as the light travel time to the BLR. This lag is traditionally used as the distance in a virial estimator, while the line widths of the variable portion of the broad lines represent the velocity component, and the virial coefficient is estimated empirically. Alternatively, forward modeling of velocity-resolved RM can provide a self-consistent mass estimate with higher precision, without relying on external calibrations of the virial coefficient \citep[e.g.,][]{Pancoast+14,Williams+18,Williams+22,Villafana+22}. An important byproduct of RM is the so-called single-epoch method, which also uses the virial theorem to estimate $\mbh$ \citep{Vestergaard+06,Vestergaard+09}; here, the BLR size ($\rblr$) is approximated from the monochromatic luminosity via the empirical BLR size - luminosity ($\rblr-L$) relation \citep{Kaspi+00,Kaspi+05,Bentz+13}. Unfortunately there are many factors of uncertainty in this methodology, such as which line width measurements best represent what is used in RM \citep{Assef+11,Denney+13,Denney+16}, the intrinsic scatter of the $\rblr-L$ relation, and the unknown virial coefficient. Thus, RM is arguably the best way to measure $\mbh$, in terms of both accuracy and versatility. 

The downsides of RM are twofold; the first is the large number of epochs required for monitoring, and the second is the signal-to-noise ratio (SNR) of the observations required to detect variability. It is much easier to obtain high-SNR spectra for bright QSOs, but in general these are less variable and therefore detecting variability requires high SNR \citep[hereafter \citetalias{Suberlak+21}]{Suberlak+21}, so long exposure times per epoch are necessary regardless of the intrinsic luminosity. In addition, the expected time lags for bright QSOs are in the range of 10$^{2-4}$ days, which needs to be multiplied by a factor of ($1+\zs$) to account for time dilation. The cost of a RM campaign is further exacerbated by the fact that the monitoring baseline needs to be at least several times longer than the observed time lag to detect the fluctuations \citep{ShenY+15a}, implying that the required monitoring periods for bright and distant QSOs are of decade-scales \citep{Kaspi+17}. 

Both shortcomings of the technique can be mitigated by observing gravitationally-lensed QSOs. Gravitational lensing occurs when a light source is located directly behind a massive object. When the background source is a point source, like a QSO, this results in multiple images, whose luminosities are usually magnified; the magnification enables us to measure both the continuum and emission line fluxes at lower observational cost, sometimes even for QSOs that have intrinsic brightnesses below the detection limit of a given telescope and instrument combination. A second characteristic of lensed QSOs is time delays; the multiple images of the same source QSO travel through different light paths depending on the configuration of the source and deflector, so a time delay arises between any image pair. If the multiple images are monitored simultaneously, one effectively obtains multiple epochs at once, thus reducing the requirements in terms of sampling and duration of a RM campaign. A recent study has successfully conducted RM for a distant gravitationally lensed QSO, effectively extending the intrinsic light curves by two years with respect to the duration of the monitoring period \citep{Williams+21}. Both effects are most relevant for distant sources, where only bright QSOs can be seen (and thus magnification will be important) and the time dilation effect is the greatest. 

Another application of lensed QSOs is that they can be used to constrain cosmological parameters. The time delays among the multiple images depend not only on the deflector-source configuration but also on the parameters of the universe. This allows one to measure the Hubble constant (H$_0$) with high precision that is completely independent of other methodologies \citep[see, e.g.,][for a recent review]{Treu+22}. This technique has the potential to help solve the so-called ``Hubble tension'', i.e. the tension between low and high redshift measurements of the Hubble Constant H$_0$ \citep{Abdalla+22}.

A key aspect in both the applications of lensed QSOs mentioned above is that the source QSOs need to be variable. Although the variable nature of QSOs has been recognized since their discovery and studied for decades \citep{Smith+63,Matthews+63}, the exact mechanism behind it is still not fully understood \citep{Hawkins02}. Fortunately, the mechanism of the QSO variability is unimportant in both cases, and only the phenomenon of variability itself is critical for monitoring purposes. Thus, an empirical description of the variability is sufficient for these applications. 

The magnitude and timescale of the variability are thought to depend on specific parameters of the QSO, such as its luminosity \citep{Giveon+99}. Unfortunately the correlations are difficult to confirm, to the extent that even the direction of the correlations are ambiguous from study to study (see discussions in \citealt{Giveon+99} and \citetalias{Suberlak+21}); biased samples and imperfect light curves used in the analyses have been cited as some of the causes. The most successful practical model so far has been the damped random walk (DRW) model \citep{Kelly+09}. \cite{MacLeod+10,MacLeod+12} have demonstrated that this model can be applied to large samples of individual QSO light curves as well, and \citetalias{Suberlak+21} present updates with longer observation baselines. Even though modifications to this simple approach could provide an even better description of individual QSO light curves \citep{Kelly+14}, DRW suffices for most applications.

In this paper, we estimate how many lensed QSOs in the Legacy Survey of Space and Time (LSST) will exhibit sufficiently large variability and luminosity to be suitable candidates for RM or measuring time delays. Section \ref{sec:method} illustrates the methodology. Results are shown in Section \ref{sec:results}, and discussed in Section \ref{sec:disc}. The paper is concluded in Section \ref{sec:conc}. Throughout this paper, we use the AB magnitude system \citep{Oke+83} and a standard $\Lambda$CDM cosmology with H$_0=70$~\kms{}~Mpc$^{-1}$, $\Omega_{\rm M}=0.3$ and $\Omega_\Lambda=0.7$, unless specified otherwise.

\section{Methodology}\label{sec:method}

We start with the lensed QSO mock catalog provided by \citet[hereafter \citetalias{OguriM+10}]{OguriM+10}. The catalog contains 15658 mock lensed QSOs over a 100,000~deg$^2$ area, and the parameters of the deflectors and sources for each lensed QSO, such as redshifts of the deflector and source, magnitudes of the source and lensed images, and time delays. For our predictions, we use some parameters that already exist in the catalog, and compute some additional parameters discussed below. Finally, to match with the expected LSST areal coverage of 20,000~deg$^2$, we scale the numbers by 1/5 for a sample of 3131.6 lensed QSOs. We note that the cosmology used for the catalog assumes $\Omega_{\rm M}=0.26$, which is different from what is used in other parts of our paper, but we do not expect this difference to be significant for our purposes. 

\subsection{Asymptotic Structure Function} \label{subsec:sf}
To model the variability, we follow the DRW model for SDSS QSOs shown in \citetalias{Suberlak+21}. From a time-series light curve of a QSO, we can calculate the structure function (SF), which is the root-mean-squared scatter of magnitude differences for a fixed time difference $\Delta t$, as an exponentially-converging function in the form of 
\begin{equation}
\text{SF}(\Delta t) = \text{SF}_\infty [1-\exp(-\Delta t/\tau)]^{1/2} \label{eq:sf}
\end{equation}
where \sfinf{} is the asymptotic SF for large $\Delta t$ and $\tau$ is a characteristic timescale. We treat \sfinf{} as the magnitude of the variability, in that if \sfinf{} is larger than the magnitude errors \citepalias[Figure~8]{Suberlak+21} we regard the QSO variability to be detectable. We obtain the predicted magnitude errors for LSST from \cite{Ivezic+19}. 

Following \citetalias{Suberlak+21}, both \sfinf{} and $\tau$ are expressed as functions of the absolute magnitude and SMBH mass of the quasar as 
%
\begin{equation}
\log f = A + B \log \dfrac{\lambda_{\rm RF}}{4000~\text{\AA}} + C (M_i+23) + D \log \dfrac{\mbh}{10^9~M_\odot},\label{eq:sftau}
\end{equation}
%
where $f$ is either \sfinf{} or $\tau$, $\lambda_{\rm RF}$ is the restframe wavelength, $M_i$ is the $i$-band absolute magnitude, $\mbh$ is the mass of the SMBH, and $A$ through $D$ are the coefficients shown in Table~2 of \citetalias{Suberlak+21}. Among the three coefficient sets in the table, we use those for the SDSS-PS1 combined light curves. 

Since $\mbh$ is not a given quantity in the \citetalias{OguriM+10} catalog, we estimate it from the QSO luminosity following \cite{TaakY+20}, and describe this procedure here concisely. For QSOs within a specific bin for the absolute magnitude in the $i$-filter for $z=2$ ($M_{i,z=2}$), the $\mbh$ distribution can be fit as a Gaussian function. The mean $\mbh$ of this Gaussian for all $M_{i,z=2}$ bins is a linear function of $M_{i,z=2}$ (Figure 4(b) of \citealt{TaakY+20}), so we can estimate the $\mbh$ of a QSO as $\log (\mbh/\msun)=2.5-0.25 M_{i,z=2}$ with errors of $\sim0.3$~dex. This estimate is in agreement with Figure~21(a) of \citetalias{Suberlak+21}, thus demonstrating the validity of the method. Since the values for coefficients C and D are comparable, this equation suggests that the overall dependence of \sfinf{} on the QSO brightness is negative, i.e., brighter QSOs are less variable, as expected. 

For two-image lens systems (doubles), the variability of the dimmer image determines whether a lensed QSO can be used for either RM or H$_0$ measurements; for four-image lens systems (quads), detection and monitoring of at least the third-brightest image is necessary to make use of information content beyond that of a double \citepalias{OguriM+10}. Thus, we introduce the ``reference'' magnitude, which is the magnitude of either the dimmer image for doubles or that of the third-brightest image for quads. 

\begin{figure}
\centering
\includegraphics[width=\textwidth]{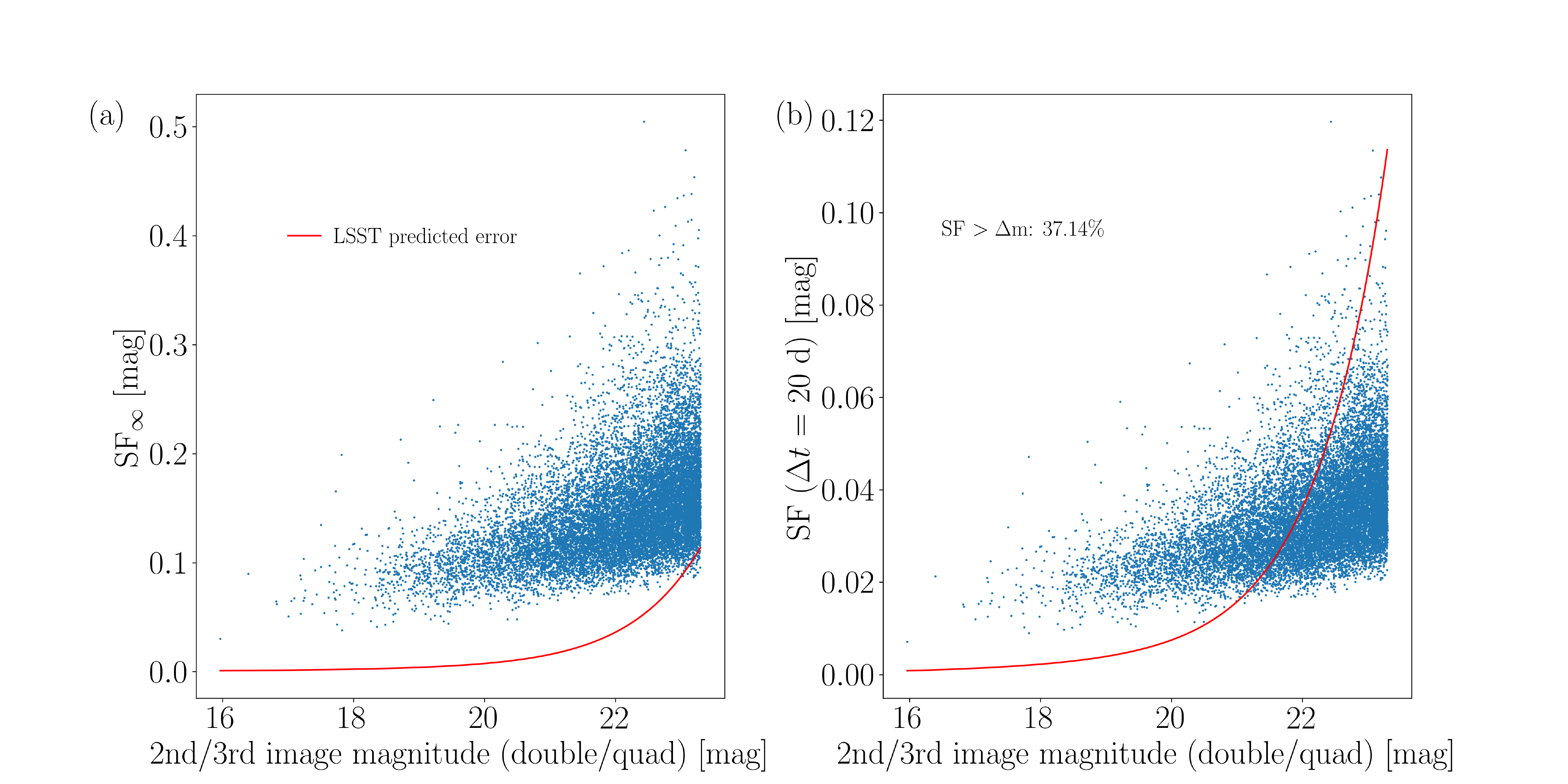}
\caption{(a) Scatter plot of \sfinf{} versus the reference magnitude for the lensed QSO sample from \citetalias{OguriM+10}. The red solid line denotes the predicted error for single-epoch LSST observations \citep{Ivezic+19}.
(b) Scatter plot of the structure function for $\Delta t$=20 days versus the reference magnitude. 
}
\label{fig:sf}
\end{figure}

Figure~\ref{fig:sf}(a) shows the scatter plot of \sfinf{} versus the reference magnitude. As expected from Eq.~\ref{eq:sftau}, lensed QSOs with fainter images have a tendency to be marginally more variable with larger scatter. Figure~\ref{fig:sf}(b) shows a similar plot with SF($\Delta t$) with $\Delta t=20$~days instead, to visualize the actual magnitude differences for a LSST-like cadence. The shape of the scatter plots are similar apart from a scaling factor, indicating that SF($\Delta t=20$~d) is simply \sfinf{} reduced by a factor of $\sim5$. We assume for simplicity that the QSO variability can be detected if SF($\Delta t$) is larger than the predicted magnitude errors ($\Delta m$) for LSST \citep[Eq.~5]{Ivezic+19}, which are also plotted for comparison. Using this argument, we estimate that variability will be detectable for 37\% of the lensed QSO population. 

\subsection{BLR size}
\begin{figure}
\centering
\includegraphics[width=\textwidth]{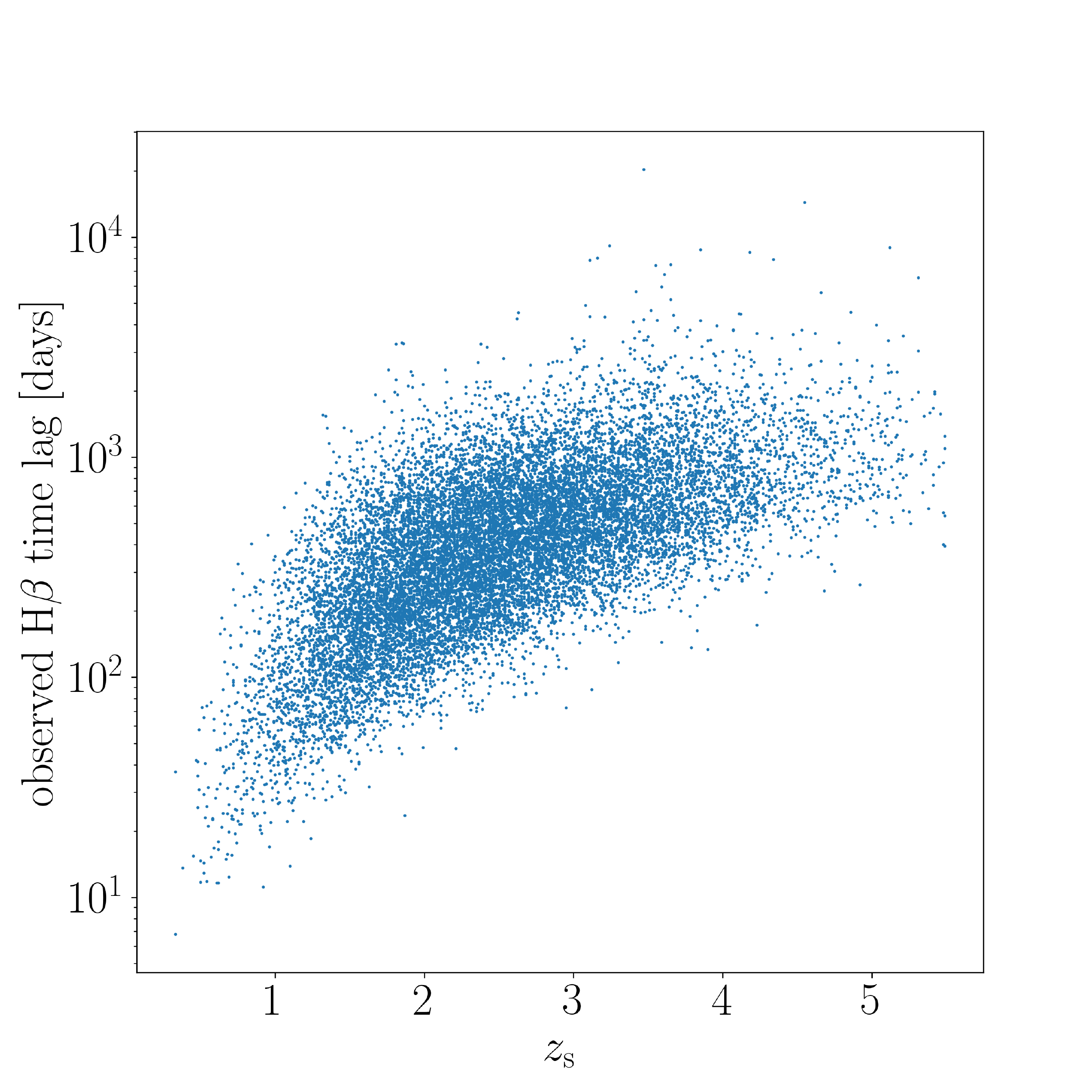}
\caption{Scatter plot of the expected observed time lag for the H$\beta$ line versus the source redshift for the lensed QSO sample from \citetalias{OguriM+10}. 
}
\label{fig:lag}
\end{figure}

For RM purposes, $\rblr$ is an important property in that it allows us to estimate the observation duration required for measuring the time lags. It can be predicted from the luminosity of the AGN using the empirical relation \citep{Bentz+13} 
\begin{equation}
\begin{aligned}
\log \dfrac{\rblr}{\text{lt-days}} = 1.527 + 0.533 \log \dfrac{L_{\rm 5100}}{\rm 10^{44}~erg~s^{-1}},
\label{eq:rblr}\\
\end{aligned}
\end{equation}
where $L_{\rm 5100}$ is the monochromatic luminosity of the AGN at 5100~\AA. We estimate the observed time lag by using this BLR size multiplied by a factor of (1+$z$) to account for time dilation. These lags are plotted versus the source redshift in Figure~\ref{fig:lag}.

\section{Results} \label{sec:results}

\begin{figure}
\centering
\includegraphics[width=\textwidth]{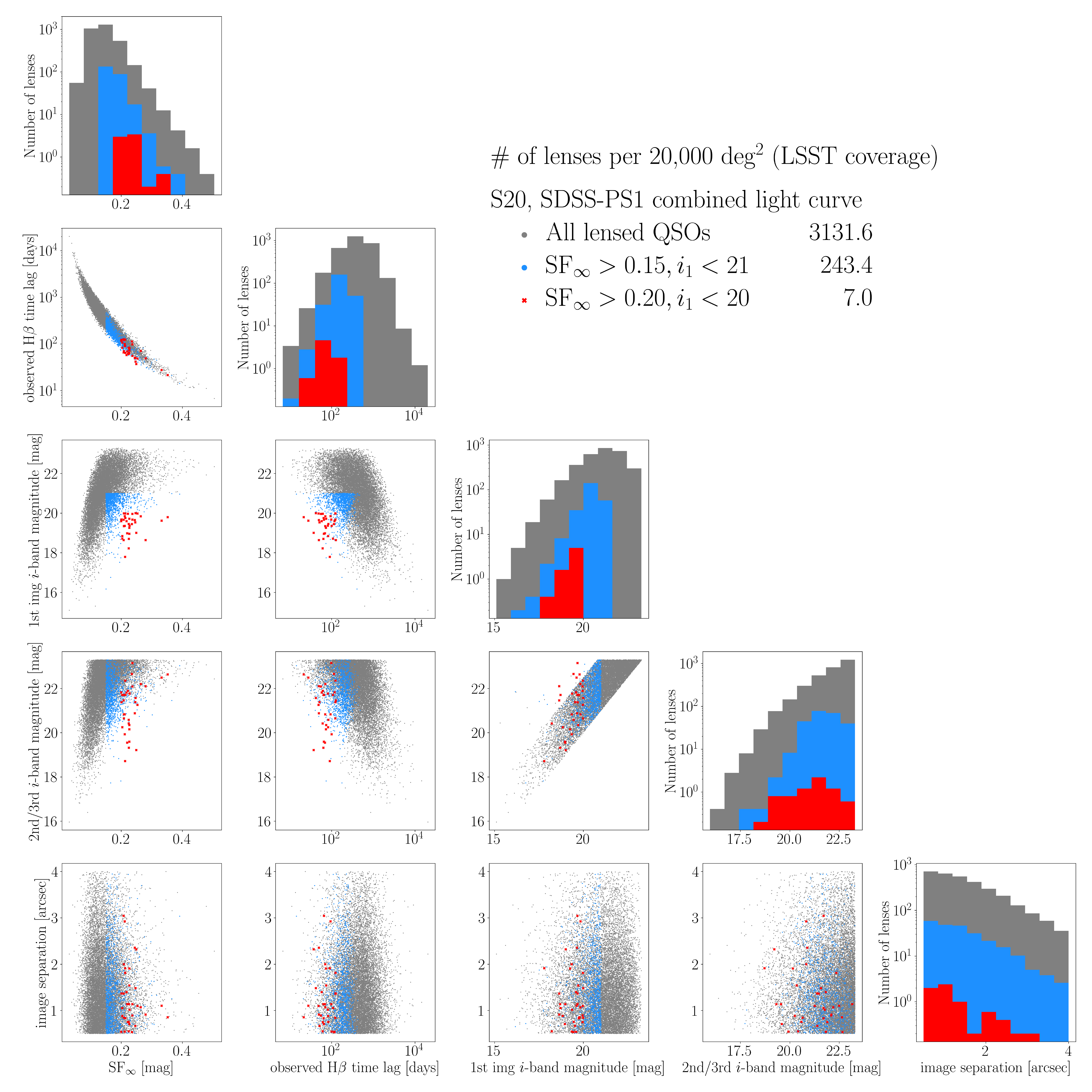}
\caption{Scatter plot of several parameters of the \citetalias{OguriM+10} lensed QSO sample. Gray dots indicate the full sample, while blue dots and red crosses represent the subsample of lensed QSOs with \sfinf{} and $i_1$ cuts. The number of lensed QSOs in each sample scaled for a fiducial LSST-like areal coverage are shown on the plot. The scatter plots show the samples for the unscaled 100,000~deg$^2$ areal coverage, while the histograms for each parameter on the diagonal show the scaled numbers. 
}
\label{fig:scatter}
\end{figure}

Figure~\ref{fig:scatter} shows the scatter plots of various properties of the lensed QSOs computed in the previous section. In the previous section we discussed the detectability of variability. In this section we focus on quantifying the size of subsamples that would be ideal for RM and time delays studies, i.e. those 
that exhibit bright lensed images with large variability. ``Bright'' and ``large'' in this context depend mostly on the observational resources that one has available, and there is certainly a degree of subjectivity in presenting any specific choice. We present some for illustration, making the full code available to the reader, in case they want to pursue their own selection \footnote{The code will be available at \url{https://github.com/yctaak/Lensed-QSO-Variability} upon acceptance of the paper.}.

The cuts applied for the first subsample (Subsample 1) are \sfinf{} > 0.15 and $i_1 < 21$~mag, where $i_1$ is the magnitude of the brightest image; these numbers represent variability scales that can be readily monitored by two- to four-meter-class telescopes \citep{ShenY+15a}. The number of lensed QSOs satisfying these conditions is 243.4 for a 20,000~deg$^2$ areal coverage. We apply a second set of cuts that chooses a smaller subsample, Subsample 2, displaying larger variability for brighter images. The cuts applied for Subsample 2 are \sfinf{} > 0.20 and $i_1 < 20$~mag, and the number of lensed QSOs in this subsample is 7.0. These numbers demonstrate that there is a significant number of lensed QSOs within the LSST areal coverage, to be followed up with two- to four-meter-class telescopes. Clearly, a follow-up with larger telescopes would have access to a much larger subsample.  For example, \citet{Williams+21} monitored a source of apparent magnitude $i\sim 20-21$ using the Gemini North 8-m Telescope, achieving velocity-resolved RM measurements. Likewise, high-cadence monitoring with milli-magnitude precision would be able to determine gravitational time delays even with small variability amplitude over a relatively short timescale \citep{Courbin+18}. 

\begin{figure}
\centering
\includegraphics[width=\textwidth]{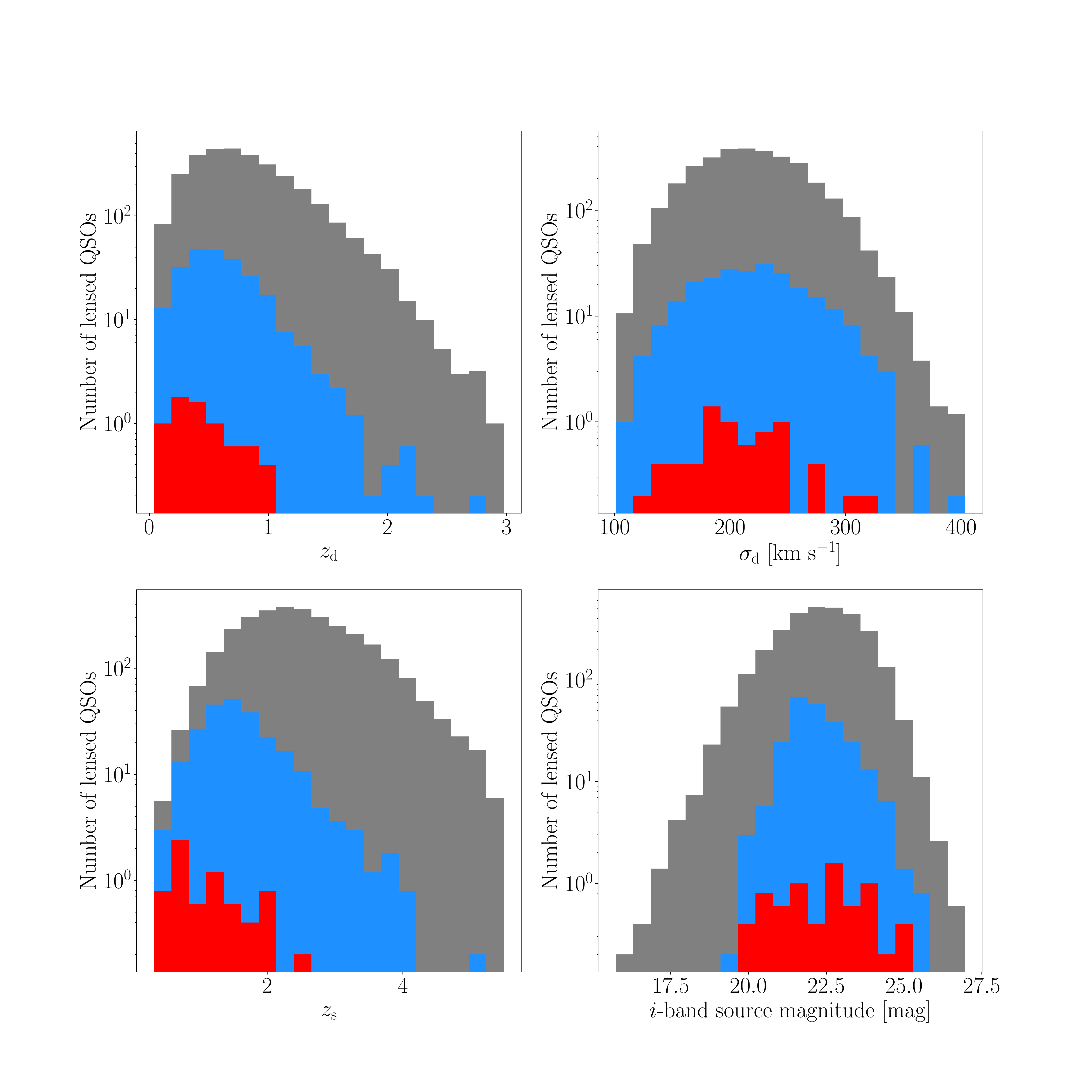}
\caption{Histogram of the deflector and source properties for the \citetalias{OguriM+10} sample. Numbers are scaled for a 20,000~deg$^2$ areal coverage. Top row: deflector redshift and deflector velocity dispersion. Bottom row: source redshift and $i$-band source magnitude. Colors are identical to those in Figure~\ref{fig:scatter}. 
}
\label{fig:zhist}
\end{figure}

The distributions of some of the lensed QSO properties for the full sample and each subsample are shown in Figure~\ref{fig:zhist}. The distributions for the deflector properties are similar between the three samples (Subsamples 1, 2, and the full sample) except for an overall scale-factor difference, suggesting that there are no selection effects due to the subsamples regarding the deflector population. We note that the number of lensed QSOs with $\zs>1$ account for at least one half of each subsample, demonstrating a plethora of distant targets for RM, which benefit the most from the use of lensed QSOs due to the effect of time dilation as was discussed in the introduction. 

\begin{figure}
\centering
\includegraphics[width=\textwidth]{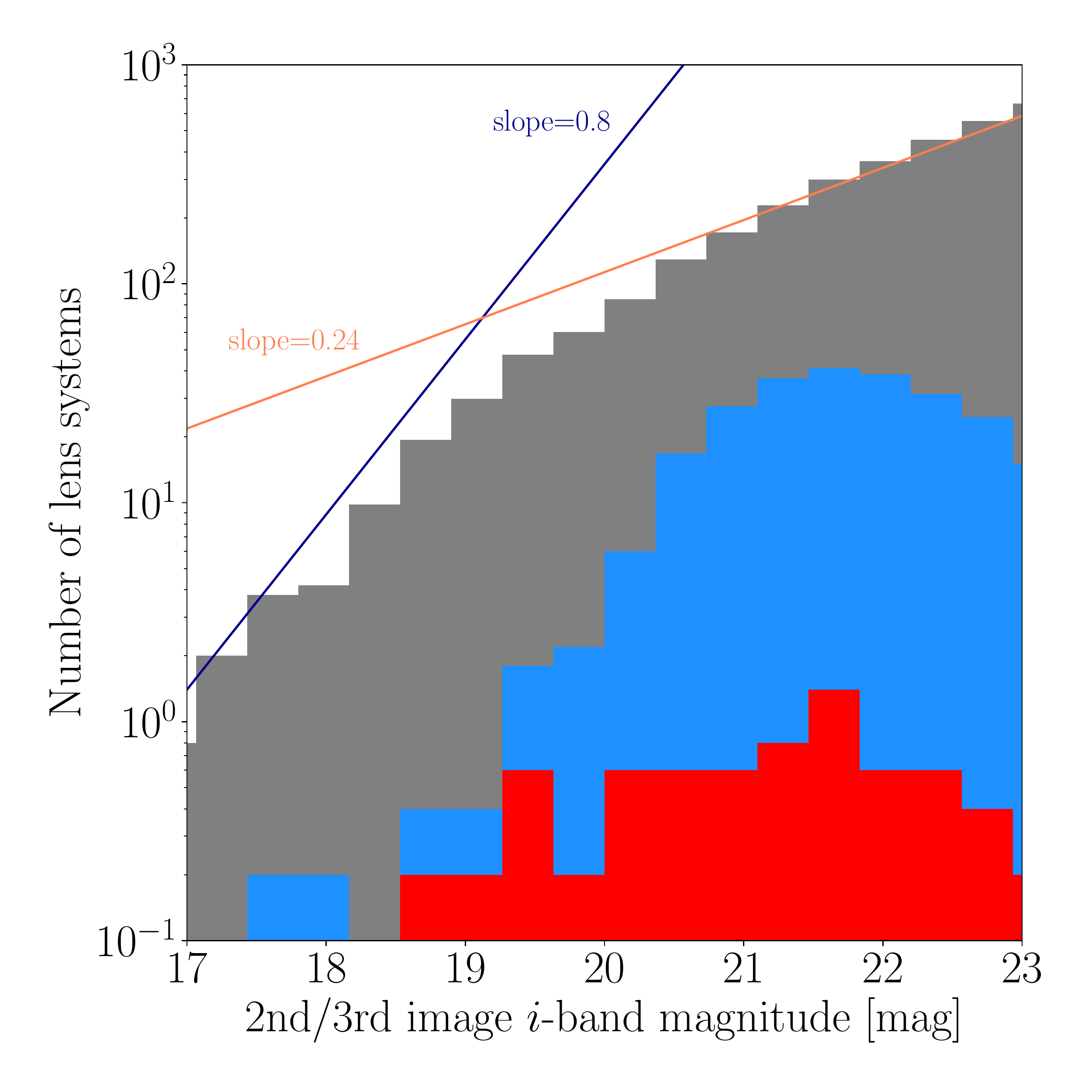}
\caption{Histogram of the reference magnitude for the three subsamples. The colors of the histograms match those used in Figure~\ref{fig:scatter}. The dark blue line indicates a slope of 0.8 corresponding to depth-width equivalence, while the pink line corresponds to the slope of the histogram at $i_{\rm ref}\sim$~22--23~mag.
}
\label{fig:depth}
\end{figure}

It is possible to determine whether area or depth is more important for these lensed QSOs. Following \citet{TaakY+20}, we consider two surveys with identical observation times, with one survey aiming for depth and the other for wider area. If the area ratio is 1:10, the deeper survey should have a limiting magnitude that is 1.25~mag deeper. So for the deeper survey to yield as many lensed QSOs as the wider survey, the slope of the magnitude histogram must be ($\log 10$)/1.25 = 0.8.  Figure~\ref{fig:depth} shows the histogram of the reference magnitude for the full sample and two subsamples. We can see that the slope of 0.8 is only achievable for shallow surveys, implying that for depths similar to those for LSST, a wider area should be prioritized over depth to maximize the lensed QSO sample size, which is what has been done for LSST.

\section{Discussion: Sources of Uncertainty} \label{sec:disc}

\subsection{Variants to \sfinf{} Computation}
Table~2 of \citetalias{Suberlak+21} provides three distinct sets of coefficients for \sfinf{} and $\tau$, depending on the light curves used for the analysis and the method used for obtaining the coefficients. The first and second sets use SDSS-only light curves which span $\sim10$~years and are used in \cite{MacLeod+10}, while the final coefficient set uses SDSS-PS1 combined light curves, which effectively extends the timeline to 15~years. The difference between the first and second coefficient sets is that the first set provides the best-fit coefficients (labeled as the M10 method) while the second set uses likelihood distributions to calculate the expectation values (labeled as the S20 method, in line with the notation in \citetalias{Suberlak+21}). The final coefficient set, which is the one used in Section~\ref{subsec:sf}, also uses the S20 method, but is derived from the best fits to the SDSS-PS1 combined light curves, which we trust to be more reliable due to its longer observation baseline. 

To test the robustness of our results with respect to the various approaches used to estimate the variability, we recalculate the number of lensed QSOs that satisfy our selection cuts using the other two coefficient sets for the SDSS light curves shown in Table~2 of \citetalias{Suberlak+21}. The number of lensed QSOs satisfying the two conditions for \sfinf{} and $i_1$ (Subsamples 1 and 2) are summarized in Table~\ref{tbl:lensedqso}. Based on these numbers, we conclude that the number of bright and variable lensed QSOs may be decreased by a factor of 2 with respect to the numbers shown in Section~\ref{sec:results}, depending on the coefficient set used for the \sfinf{} computation. We can see that the sample sizes for both subsamples decrease in the following order: S20/SDSS-PS1, M10/SDSS, S20/SDSS. The main reason for this difference is the change in the offset of the coefficient sets, A, for \sfinf{}; the values of A decrease in the same order, meaning that \sfinf{} becomes smaller in the same order, and consequently the number of lensed QSOs satisfying the subsample cut decreases. The fraction of all lensed QSOs that have SF($\Delta t$) larger than its predicted magnitude error for LSST-like observations changes from 37\% for the coefficient set using the SDSS-PS1 combined light curve to 41\% and 34\% for the two coefficient sets using the SDSS-only light curve. 

\subsection{Updated Lensed QSO Mock Catalog}
The \citetalias{OguriM+10} catalog uses a deflector velocity dispersion function (VDF) that is constant over cosmic time, along with an outdated version of the redshift-dependent QSO luminosity function (LF). \citet{YueM+22} have updated this catalog in three major aspects; first by employing a revised VDF from the CosmoDC2 mock galaxy catalog \citep{Korytov+19} which allows for a redshift evolution of the deflector VDF, second by using a QSO LF which includes the latest discoveries especially at higher redshifts, and third by extending the image separation limit down to $\sim0\farcs1$. We select lensed QSOs with reference magnitudes brighter than the LSST single-visit depth of $i=23.3$~mag and image separations larger than $0\farcs5$ for direct comparison with the \citetalias{OguriM+10} catalog. For the \sfinf{} calculation, we use only the S20 method for the SDSS-PS1 combined light curve for comparison purposes. The number of lensed QSOs are scaled to match the fiducial areal coverage of 20,000~deg$^2$.

The results are as follows; the number of lensed QSOs for a LSST-like areal coverage is 813.0, and the sizes of Subsamples 1 and 2 are 42.5 and 1.0, respectively. The (sub)sample sizes are smaller than those for the \citetalias{OguriM+10} catalog by a factor of $\sim$~4--7, which can be explained by the differences in the VDF and QSO LF used for the two catalogs. 

\subsection{Range of $\rein$}
The \citetalias{OguriM+10} catalog contains galaxy-scale deflectors only, with limits to the image separation at $0\farcs5$ and 4$\arcsec$. Lensed QSOs with larger image separations for group or cluster-scale deflectors will be especially useful for RM studies due to their longer and thus easier-to-measure time delays. As discussed earlier, these longer time delays will be most effective for the reduction of the monitoring baseline.  Therefore we should consider our estimate conservative in this regard, since systems with wider image separation should be discoverable within LSST, as they have been discovered in SDSS.

At the other end of the spectrum, the distribution of the Einstein radii of galaxy-scale deflectors \citep{Collett15,TaakY+20} suggests that there is a plethora of lensed QSOs with smaller image separations. Unfortunately these small-separation lensed QSOs will be difficult to monitor with ground-based telescopes such as LSST due to atmospheric seeing. Thus, our estimates are conservative in this respect as well.

\section{Conclusion} \label{sec:conc}

In this paper, we demonstrate the methodology for computing the number of lensed QSOs that are sufficiently bright and variable for RM and H$_0$ measurements with LSST. Among the 3131.6 lensed QSOs in the \citetalias{OguriM+10} mock catalog, 30--40\% will show variability larger than the LSST magnitude errors, and 100--250 of them are bright and significantly variable. An updated version of the mock catalog suggests a factor of $\sim$~5 decrease in these numbers. These numbers indicate that plenty of lensed QSOs exhibit sufficient variability for monitoring with small telescopes, and many more will be monitored systematically with LSST. The calculations performed in this paper, and the software that is publicly released with it, will help designing future studies for either RM or time delay measurements.

\section*{Acknowledgements}

We thank Masamune Oguri, Philip J. Marshall, Minghao Yue, and Matt Malkan for providing valuable discussions. This research is supported by the Basic Science Research Program through the National Research Foundation of Korea (NRF) funded by the Ministry of Education (grant number 2021R1A6A3A14044070).

\section*{Data Availability}

The code for the subsample selection process, along with its resulting catalog, will be available at \url{https://github.com/yctaak/Lensed-QSO-Variability} upon acceptance of the paper.



\bibliographystyle{mnras}
\bibliography{bibtex} 

\begin{thebibliography}{}
\makeatletter
\relax
\def\mn@urlcharsother{\let\do\@makeother \do\$\do\&\do\#\do\^\do\_\do\%\do\~}
\def\mn@doi{\begingroup\mn@urlcharsother \@ifnextchar [ {\mn@doi@}
  {\mn@doi@[]}}
\def\mn@doi@[#1]#2{\def\@tempa{#1}\ifx\@tempa\@empty \href
  {http://dx.doi.org/#2} {doi:#2}\else \href {http://dx.doi.org/#2} {#1}\fi
  \endgroup}
\def\mn@eprint#1#2{\mn@eprint@#1:#2::\@nil}
\def\mn@eprint@arXiv#1{\href {http://arxiv.org/abs/#1} {{\tt arXiv:#1}}}
\def\mn@eprint@dblp#1{\href {http://dblp.uni-trier.de/rec/bibtex/#1.xml}
  {dblp:#1}}
\def\mn@eprint@#1:#2:#3:#4\@nil{\def\@tempa {#1}\def\@tempb {#2}\def\@tempc
  {#3}\ifx \@tempc \@empty \let \@tempc \@tempb \let \@tempb \@tempa \fi \ifx
  \@tempb \@empty \def\@tempb {arXiv}\fi \@ifundefined
  {mn@eprint@\@tempb}{\@tempb:\@tempc}{\expandafter \expandafter \csname
  mn@eprint@\@tempb\endcsname \expandafter{\@tempc}}}

\bibitem[\protect\citeauthoryear{{Abdalla} et~al.,}{{Abdalla}
  et~al.}{2022}]{Abdalla+22}
{Abdalla} E.,  et~al., 2022, \mn@doi [Journal of High Energy Astrophysics]
  {10.1016/j.jheap.2022.04.002}, \href
  {https://ui.adsabs.harvard.edu/abs/2022JHEAp..34...49A} {34, 49}

\bibitem[\protect\citeauthoryear{{Assef} et~al.,}{{Assef}
  et~al.}{2011}]{Assef+11}
{Assef} R.~J.,  et~al., 2011, \mn@doi [\apj] {10.1088/0004-637X/742/2/93},
  \href {https://ui.adsabs.harvard.edu/abs/2011ApJ...742...93A} {742, 93}

\bibitem[\protect\citeauthoryear{{Ba{\~n}ados} et~al.,}{{Ba{\~n}ados}
  et~al.}{2016}]{Banados+16}
{Ba{\~n}ados} E.,  et~al., 2016, \mn@doi [\apjs] {10.3847/0067-0049/227/1/11},
  \href {https://ui.adsabs.harvard.edu/abs/2016ApJS..227...11B} {227, 11}

\bibitem[\protect\citeauthoryear{{Bentz} et~al.,}{{Bentz}
  et~al.}{2013}]{Bentz+13}
{Bentz} M.~C.,  et~al., 2013, \mn@doi [\apj] {10.1088/0004-637X/767/2/149},
  \href {https://ui.adsabs.harvard.edu/abs/2013ApJ...767..149B} {767, 149}

\bibitem[\protect\citeauthoryear{{Blandford} \& {McKee}}{{Blandford} \&
  {McKee}}{1982}]{Blandford+82}
{Blandford} R.~D.,  {McKee} C.~F.,  1982, \mn@doi [\apj] {10.1086/159843},
  \href {https://ui.adsabs.harvard.edu/abs/1982ApJ...255..419B} {255, 419}

\bibitem[\protect\citeauthoryear{{Cappellari}, {Neumayer}, {Reunanen}, {van der
  Werf}, {de Zeeuw}  \& {Rix}}{{Cappellari} et~al.}{2009}]{Cappellari+09}
{Cappellari} M.,  {Neumayer} N.,  {Reunanen} J.,  {van der Werf} P.~P.,  {de
  Zeeuw} P.~T.,   {Rix} H.~W.,  2009, \mn@doi [\mnras]
  {10.1111/j.1365-2966.2008.14377.x}, \href
  {https://ui.adsabs.harvard.edu/abs/2009MNRAS.394..660C} {394, 660}

\bibitem[\protect\citeauthoryear{{Collett}}{{Collett}}{2015}]{Collett15}
{Collett} T.~E.,  2015, \mn@doi [\apj] {10.1088/0004-637X/811/1/20}, \href
  {https://ui.adsabs.harvard.edu/abs/2015ApJ...811...20C} {811, 20}

\bibitem[\protect\citeauthoryear{{Courbin} et~al.,}{{Courbin}
  et~al.}{2018}]{Courbin+18}
{Courbin} F.,  et~al., 2018, \mn@doi [\aap] {10.1051/0004-6361/201731461},
  \href {https://ui.adsabs.harvard.edu/abs/2018A&A...609A..71C} {609, A71}

\bibitem[\protect\citeauthoryear{{Denney}, {Pogge}, {Assef}, {Kochanek},
  {Peterson}  \& {Vestergaard}}{{Denney} et~al.}{2013}]{Denney+13}
{Denney} K.~D.,  {Pogge} R.~W.,  {Assef} R.~J.,  {Kochanek} C.~S.,  {Peterson}
  B.~M.,   {Vestergaard} M.,  2013, \mn@doi [\apj]
  {10.1088/0004-637X/775/1/60}, \href
  {https://ui.adsabs.harvard.edu/abs/2013ApJ...775...60D} {775, 60}

\bibitem[\protect\citeauthoryear{{Denney} et~al.,}{{Denney}
  et~al.}{2016}]{Denney+16}
{Denney} K.~D.,  et~al., 2016, \mn@doi [\apjs] {10.3847/0067-0049/224/2/14},
  \href {https://ui.adsabs.harvard.edu/abs/2016ApJS..224...14D} {224, 14}

\bibitem[\protect\citeauthoryear{{Ferrarese} \& {Merritt}}{{Ferrarese} \&
  {Merritt}}{2000}]{Ferrarese+00}
{Ferrarese} L.,  {Merritt} D.,  2000, \mn@doi [\apjl] {10.1086/312838}, \href
  {https://ui.adsabs.harvard.edu/abs/2000ApJ...539L...9F} {539, L9}

\bibitem[\protect\citeauthoryear{{Fromerth} \& {Melia}}{{Fromerth} \&
  {Melia}}{2000}]{Fromerth+00}
{Fromerth} M.~J.,  {Melia} F.,  2000, \mn@doi [\apj] {10.1086/308671}, \href
  {https://ui.adsabs.harvard.edu/abs/2000ApJ...533..172F} {533, 172}

\bibitem[\protect\citeauthoryear{{Gebhardt} et~al.,}{{Gebhardt}
  et~al.}{2000}]{Gebhardt+00}
{Gebhardt} K.,  et~al., 2000, \mn@doi [\apjl] {10.1086/312840}, \href
  {https://ui.adsabs.harvard.edu/abs/2000ApJ...539L..13G} {539, L13}

\bibitem[\protect\citeauthoryear{{Ghez} et~al.,}{{Ghez} et~al.}{2008}]{Ghez+08}
{Ghez} A.~M.,  et~al., 2008, \mn@doi [\apj] {10.1086/592738}, \href
  {https://ui.adsabs.harvard.edu/abs/2008ApJ...689.1044G} {689, 1044}

\bibitem[\protect\citeauthoryear{{Giveon}, {Maoz}, {Kaspi}, {Netzer}  \&
  {Smith}}{{Giveon} et~al.}{1999}]{Giveon+99}
{Giveon} U.,  {Maoz} D.,  {Kaspi} S.,  {Netzer} H.,   {Smith} P.~S.,  1999,
  \mn@doi [\mnras] {10.1046/j.1365-8711.1999.02556.x}, \href
  {https://ui.adsabs.harvard.edu/abs/1999MNRAS.306..637G} {306, 637}

\bibitem[\protect\citeauthoryear{{Hawkins}}{{Hawkins}}{2002}]{Hawkins02}
{Hawkins} M.~R.~S.,  2002, \mn@doi [\mnras] {10.1046/j.1365-8711.2002.04939.x},
  \href {https://ui.adsabs.harvard.edu/abs/2002MNRAS.329...76H} {329, 76}

\bibitem[\protect\citeauthoryear{{Ivezi{\'c}} et~al.,}{{Ivezi{\'c}}
  et~al.}{2019}]{Ivezic+19}
{Ivezi{\'c}} {\v{Z}}.,  et~al., 2019, \mn@doi [\apj]
  {10.3847/1538-4357/ab042c}, \href
  {https://ui.adsabs.harvard.edu/abs/2019ApJ...873..111I} {873, 111}

\bibitem[\protect\citeauthoryear{{Jeon}, {Im}, {Pak}, {Hyun}, {Kim}, {Kim},
  {Lee}  \& {Park}}{{Jeon} et~al.}{2016}]{JeonY+16}
{Jeon} Y.,  {Im} M.,  {Pak} S.,  {Hyun} M.,  {Kim} S.,  {Kim} Y.,  {Lee} H.-I.,
    {Park} W.,  2016, \mn@doi [Journal of Korean Astronomical Society]
  {10.5303/JKAS.2016.49.1.25}, \href
  {https://ui.adsabs.harvard.edu/abs/2016JKAS...49...25J} {49, 25}

\bibitem[\protect\citeauthoryear{{Jeon} et~al.,}{{Jeon}
  et~al.}{2017}]{JeonY+17}
{Jeon} Y.,  et~al., 2017, \mn@doi [\apjs] {10.3847/1538-4365/aa7de5}, \href
  {https://ui.adsabs.harvard.edu/abs/2017ApJS..231...16J} {231, 16}

\bibitem[\protect\citeauthoryear{{Jiang} et~al.,}{{Jiang}
  et~al.}{2016}]{JiangL+16}
{Jiang} L.,  et~al., 2016, \mn@doi [\apj] {10.3847/1538-4357/833/2/222}, \href
  {https://ui.adsabs.harvard.edu/abs/2016ApJ...833..222J} {833, 222}

\bibitem[\protect\citeauthoryear{{Kaspi}, {Smith}, {Netzer}, {Maoz}, {Jannuzi}
  \& {Giveon}}{{Kaspi} et~al.}{2000}]{Kaspi+00}
{Kaspi} S.,  {Smith} P.~S.,  {Netzer} H.,  {Maoz} D.,  {Jannuzi} B.~T.,
  {Giveon} U.,  2000, \mn@doi [\apj] {10.1086/308704}, \href
  {https://ui.adsabs.harvard.edu/abs/2000ApJ...533..631K} {533, 631}

\bibitem[\protect\citeauthoryear{{Kaspi}, {Maoz}, {Netzer}, {Peterson},
  {Vestergaard}  \& {Jannuzi}}{{Kaspi} et~al.}{2005}]{Kaspi+05}
{Kaspi} S.,  {Maoz} D.,  {Netzer} H.,  {Peterson} B.~M.,  {Vestergaard} M.,
  {Jannuzi} B.~T.,  2005, \mn@doi [\apj] {10.1086/431275}, \href
  {https://ui.adsabs.harvard.edu/abs/2005ApJ...629...61K} {629, 61}

\bibitem[\protect\citeauthoryear{{Kaspi}, {Brandt}, {Maoz}, {Netzer},
  {Schneider}  \& {Shemmer}}{{Kaspi} et~al.}{2017}]{Kaspi+17}
{Kaspi} S.,  {Brandt} W.~N.,  {Maoz} D.,  {Netzer} H.,  {Schneider} D.~P.,
  {Shemmer} O.,  2017, \mn@doi [Frontiers in Astronomy and Space Sciences]
  {10.3389/fspas.2017.00031}, \href
  {https://ui.adsabs.harvard.edu/abs/2017FrASS...4...31K} {4, 31}

\bibitem[\protect\citeauthoryear{{Kelly}, {Bechtold}  \&
  {Siemiginowska}}{{Kelly} et~al.}{2009}]{Kelly+09}
{Kelly} B.~C.,  {Bechtold} J.,   {Siemiginowska} A.,  2009, \mn@doi [\apj]
  {10.1088/0004-637X/698/1/895}, \href
  {https://ui.adsabs.harvard.edu/abs/2009ApJ...698..895K} {698, 895}

\bibitem[\protect\citeauthoryear{{Kelly}, {Becker}, {Sobolewska},
  {Siemiginowska}  \& {Uttley}}{{Kelly} et~al.}{2014}]{Kelly+14}
{Kelly} B.~C.,  {Becker} A.~C.,  {Sobolewska} M.,  {Siemiginowska} A.,
  {Uttley} P.,  2014, \mn@doi [\apj] {10.1088/0004-637X/788/1/33}, \href
  {https://ui.adsabs.harvard.edu/abs/2014ApJ...788...33K} {788, 33}

\bibitem[\protect\citeauthoryear{{Kim} et~al.,}{{Kim} et~al.}{2015}]{KimY+15}
{Kim} Y.,  et~al., 2015, \mn@doi [\apjl] {10.1088/2041-8205/813/2/L35}, \href
  {https://ui.adsabs.harvard.edu/abs/2015ApJ...813L..35K} {813, L35}

\bibitem[\protect\citeauthoryear{{Kim} et~al.,}{{Kim} et~al.}{2019}]{KimY+19}
{Kim} Y.,  et~al., 2019, \mn@doi [\apj] {10.3847/1538-4357/aaf387}, \href
  {https://ui.adsabs.harvard.edu/abs/2019ApJ...870...86K} {870, 86}

\bibitem[\protect\citeauthoryear{{Kim} et~al.,}{{Kim} et~al.}{2020}]{KimY+20}
{Kim} Y.,  et~al., 2020, \mn@doi [\apj] {10.3847/1538-4357/abc0ea}, \href
  {https://ui.adsabs.harvard.edu/abs/2020ApJ...904..111K} {904, 111}

\bibitem[\protect\citeauthoryear{{Kim} et~al.,}{{Kim} et~al.}{2022}]{KimY+22}
{Kim} Y.,  et~al., 2022, \mn@doi [\aj] {10.3847/1538-3881/ac81c8}, \href
  {https://ui.adsabs.harvard.edu/abs/2022AJ....164..114K} {164, 114}

\bibitem[\protect\citeauthoryear{{Kormendy} \& {Ho}}{{Kormendy} \&
  {Ho}}{2013}]{Kormendy+13}
{Kormendy} J.,  {Ho} L.~C.,  2013, \mn@doi [\araa]
  {10.1146/annurev-astro-082708-101811}, \href
  {https://ui.adsabs.harvard.edu/abs/2013ARA&A..51..511K} {51, 511}

\bibitem[\protect\citeauthoryear{{Korytov} et~al.,}{{Korytov}
  et~al.}{2019}]{Korytov+19}
{Korytov} D.,  et~al., 2019, \mn@doi [\apjs] {10.3847/1538-4365/ab510c}, \href
  {https://ui.adsabs.harvard.edu/abs/2019ApJS..245...26K} {245, 26}

\bibitem[\protect\citeauthoryear{{MacLeod} et~al.,}{{MacLeod}
  et~al.}{2010}]{MacLeod+10}
{MacLeod} C.~L.,  et~al., 2010, \mn@doi [\apj] {10.1088/0004-637X/721/2/1014},
  \href {https://ui.adsabs.harvard.edu/abs/2010ApJ...721.1014M} {721, 1014}

\bibitem[\protect\citeauthoryear{{MacLeod} et~al.,}{{MacLeod}
  et~al.}{2012}]{MacLeod+12}
{MacLeod} C.~L.,  et~al., 2012, \mn@doi [\apj] {10.1088/0004-637X/753/2/106},
  \href {https://ui.adsabs.harvard.edu/abs/2012ApJ...753..106M} {753, 106}

\bibitem[\protect\citeauthoryear{{Matthews} \& {Sandage}}{{Matthews} \&
  {Sandage}}{1963}]{Matthews+63}
{Matthews} T.~A.,  {Sandage} A.~R.,  1963, \mn@doi [\apj] {10.1086/147615},
  \href {https://ui.adsabs.harvard.edu/abs/1963ApJ...138...30M} {138, 30}

\bibitem[\protect\citeauthoryear{{Oguri} \& {Marshall}}{{Oguri} \&
  {Marshall}}{2010}]{OguriM+10}
{Oguri} M.,  {Marshall} P.~J.,  2010, \mn@doi [\mnras]
  {10.1111/j.1365-2966.2010.16639.x}, \href
  {https://ui.adsabs.harvard.edu/abs/2010MNRAS.405.2579O} {405, 2579}

\bibitem[\protect\citeauthoryear{{Oke} \& {Gunn}}{{Oke} \&
  {Gunn}}{1983}]{Oke+83}
{Oke} J.~B.,  {Gunn} J.~E.,  1983, \mn@doi [\apj] {10.1086/160817}, \href
  {https://ui.adsabs.harvard.edu/abs/1983ApJ...266..713O} {266, 713}

\bibitem[\protect\citeauthoryear{{Pancoast}, {Brewer}  \& {Treu}}{{Pancoast}
  et~al.}{2014}]{Pancoast+14}
{Pancoast} A.,  {Brewer} B.~J.,   {Treu} T.,  2014, \mn@doi [\mnras]
  {10.1093/mnras/stu1809}, \href
  {https://ui.adsabs.harvard.edu/abs/2014MNRAS.445.3055P} {445, 3055}

\bibitem[\protect\citeauthoryear{{Peterson} et~al.,}{{Peterson}
  et~al.}{2004}]{Peterson+04}
{Peterson} B.~M.,  et~al., 2004, \mn@doi [\apj] {10.1086/423269}, \href
  {https://ui.adsabs.harvard.edu/abs/2004ApJ...613..682P} {613, 682}

\bibitem[\protect\citeauthoryear{{Shen} et~al.,}{{Shen}
  et~al.}{2015a}]{ShenY+15a}
{Shen} Y.,  et~al., 2015a, \mn@doi [\apjs] {10.1088/0067-0049/216/1/4}, \href
  {https://ui.adsabs.harvard.edu/abs/2015ApJS..216....4S} {216, 4}

\bibitem[\protect\citeauthoryear{{Shen} et~al.,}{{Shen}
  et~al.}{2015b}]{ShenY+15b}
{Shen} Y.,  et~al., 2015b, \mn@doi [\apj] {10.1088/0004-637X/805/2/96}, \href
  {https://ui.adsabs.harvard.edu/abs/2015ApJ...805...96S} {805, 96}

\bibitem[\protect\citeauthoryear{{Smith} \& {Hoffleit}}{{Smith} \&
  {Hoffleit}}{1963}]{Smith+63}
{Smith} H.~J.,  {Hoffleit} D.,  1963, \mn@doi [\nat] {10.1038/198650a0}, \href
  {https://ui.adsabs.harvard.edu/abs/1963Natur.198..650S} {198, 650}

\bibitem[\protect\citeauthoryear{{Suberlak}, {Ivezi{\'c}}  \&
  {MacLeod}}{{Suberlak} et~al.}{2021}]{Suberlak+21}
{Suberlak} K.~L.,  {Ivezi{\'c}} {\v{Z}}.,   {MacLeod} C.,  2021, \mn@doi [\apj]
  {10.3847/1538-4357/abc698}, \href
  {https://ui.adsabs.harvard.edu/abs/2021ApJ...907...96S} {907, 96}

\bibitem[\protect\citeauthoryear{{Taak} \& {Im}}{{Taak} \&
  {Im}}{2020}]{TaakY+20}
{Taak} Y.~C.,  {Im} M.,  2020, \mn@doi [\apj] {10.3847/1538-4357/ab9b23}, \href
  {https://ui.adsabs.harvard.edu/abs/2020ApJ...897..163T} {897, 163}

\bibitem[\protect\citeauthoryear{{Treu}, {Suyu}  \& {Marshall}}{{Treu}
  et~al.}{2022}]{Treu+22}
{Treu} T.,  {Suyu} S.~H.,   {Marshall} P.~J.,  2022, \mn@doi [\aapr]
  {10.1007/s00159-022-00145-y}, \href
  {https://ui.adsabs.harvard.edu/abs/2022A&ARv..30....8T} {30, 8}

\bibitem[\protect\citeauthoryear{{Vestergaard} \& {Osmer}}{{Vestergaard} \&
  {Osmer}}{2009}]{Vestergaard+09}
{Vestergaard} M.,  {Osmer} P.~S.,  2009, \mn@doi [\apj]
  {10.1088/0004-637X/699/1/800}, \href
  {https://ui.adsabs.harvard.edu/abs/2009ApJ...699..800V} {699, 800}

\bibitem[\protect\citeauthoryear{{Vestergaard} \& {Peterson}}{{Vestergaard} \&
  {Peterson}}{2006}]{Vestergaard+06}
{Vestergaard} M.,  {Peterson} B.~M.,  2006, \mn@doi [\apj] {10.1086/500572},
  \href {https://ui.adsabs.harvard.edu/abs/2006ApJ...641..689V} {641, 689}

\bibitem[\protect\citeauthoryear{{Villafa{\~n}a} et~al.,}{{Villafa{\~n}a}
  et~al.}{2022}]{Villafana+22}
{Villafa{\~n}a} L.,  et~al., 2022, \mn@doi [\apj] {10.3847/1538-4357/ac6171},
  \href {https://ui.adsabs.harvard.edu/abs/2022ApJ...930...52V} {930, 52}

\bibitem[\protect\citeauthoryear{{Wang} et~al.,}{{Wang}
  et~al.}{2021}]{WangF+21}
{Wang} F.,  et~al., 2021, \mn@doi [\apjl] {10.3847/2041-8213/abd8c6}, \href
  {https://ui.adsabs.harvard.edu/abs/2021ApJ...907L...1W} {907, L1}

\bibitem[\protect\citeauthoryear{{Williams} \& {Treu}}{{Williams} \&
  {Treu}}{2022}]{Williams+22}
{Williams} P.~R.,  {Treu} T.,  2022, \mn@doi [\apj] {10.3847/1538-4357/ac8164},
  \href {https://ui.adsabs.harvard.edu/abs/2022ApJ...935..128W} {935, 128}

\bibitem[\protect\citeauthoryear{{Williams} et~al.,}{{Williams}
  et~al.}{2018}]{Williams+18}
{Williams} P.~R.,  et~al., 2018, \mn@doi [\apj] {10.3847/1538-4357/aae086},
  \href {https://ui.adsabs.harvard.edu/abs/2018ApJ...866...75W} {866, 75}

\bibitem[\protect\citeauthoryear{{Williams} et~al.,}{{Williams}
  et~al.}{2021}]{Williams+21}
{Williams} P.~R.,  et~al., 2021, \mn@doi [\apj] {10.3847/1538-4357/abe943},
  \href {https://ui.adsabs.harvard.edu/abs/2021ApJ...911...64W} {911, 64}

\bibitem[\protect\citeauthoryear{{Yue}, {Fan}, {Yang}  \& {Wang}}{{Yue}
  et~al.}{2022}]{YueM+22}
{Yue} M.,  {Fan} X.,  {Yang} J.,   {Wang} F.,  2022, \mn@doi [\aj]
  {10.3847/1538-3881/ac4cb0}, \href
  {https://ui.adsabs.harvard.edu/abs/2022AJ....163..139Y} {163, 139}

\makeatother
\end{thebibliography}

\clearpage

\clearpage

\clearpage

\clearpage

\clearpage

\clearpage

\begin{table}
\caption{Number of lensed QSOs for a LSST-like areal coverage with bright images and large variability}
\label{tbl:lensedqso}
\centering
\begin{tabular}{l c c c}
\hline\hline
Method$^a$ \& light curve  & \sfinf{}$ > 0.15, i_1 < 21$~mag & \sfinf{}$ > 0.2, i_1 < 20$~mag & Fraction of lensed QSOs with \\
& (Subsample 1) & (Subsample 2) & SF($\Delta t=20$~d) > $\Delta m$ \\
\hline
S20/SDSS-PS1 & 243.4 & 7.0 & 0.371\\
M10/SDSS & 211.0 & 6.2 & 0.407\\
S20/SDSS & 105.2 & 3.8 & 0.343\\
\hline
\end{tabular}
\end{table}
\footnotesize{$^a$ S20 denotes the method used in \cite{Suberlak+21}, in line with the notation within the paper. M10 denotes the method used in \cite{MacLeod+10}.}\\

\bsp	
\label{lastpage}
\end{document}